\documentclass[journal,10pt]{IEEEtran}
\usepackage{amsmath}
\usepackage{amssymb}
\usepackage{amsfonts}
\usepackage{graphicx}
\usepackage{epsfig}
\usepackage{psfrag}
\usepackage{tabulary}
\usepackage{booktabs}
\usepackage{pifont}
\usepackage{color}
\usepackage{bm,array}
\usepackage{caption}
\usepackage{url}
\usepackage[colorlinks,linkcolor=blue,anchorcolor=blue,citecolor=blue,urlcolor=black]{hyperref}
\usepackage{lettrine}
\usepackage{cite}
\usepackage{epstopdf}
\usepackage{mathtools}
\usepackage{multirow}
\usepackage{lipsum}
\usepackage{subcaption}
\usepackage{booktabs}
\usepackage{tablefootnote}

\usepackage{enumitem}

\def\({\left(}
\def\){\right)}

\def\b0{{\mathbf{0}}}

\def\mE{{\mathbb{E}}}

\def\cO{\mathcal{O}}

\newcommand{\inm}{\in\mathcal}
\newcommand{\eis}{E_i^{{(C,S)}}}
\newcommand{\ejr}{E_j^{{(C,R)}}}
\newcommand{\ris}{D_i^{{(S)}}}
\newcommand{\rir}{D_i^{{(R)}}}
\newlength{\arrow}
\newcolumntype{L}[1]{>{\raggedright\let\newline\\\arraybackslash\hspace{0pt}}m{#1}}
\newcolumntype{C}[1]{>{\centering\let\newline\\\arraybackslash\hspace{0pt}}m{#1}}
\newcolumntype{R}[1]{>{\raggedleft\let\newline\\\arraybackslash\hspace{0pt}}m{#1}}

\graphicspath{{Fig/}}
\DeclareGraphicsExtensions{.eps,.pdf,.jpeg,.png,.jpg}

\begin{document}

\title{Optimal Pricing and Load Sharing  for Energy Saving  with Cooperative Communications}
%

\author{Yinghao Guo, Lingjie Duan, and Rui Zhang
\thanks{Y. Guo is with the Department of Electrical and Computer Engineering, National University of Singapore (e-mail: yinghao.guo@nus.edu.sg).}
\thanks{L. Duan is with the Engineering Systems and Design Pillar, Singapore University of Technology and Design (e-mail: lingjie\_duan@sutd.edu.sg).}
\thanks{R. Zhang is with the Department of Electrical and Computer Engineering, National University of Singapore (e-mail:elezhang@nus.edu.sg). He is also with the Institute for Infocomm Research, A*STAR, Singapore.}}

\maketitle\thispagestyle{empty}

\newtheorem{definition}{\underline{Definition}}[section]
\newtheorem{fact}{Fact}
\newtheorem{assumption}{Assumption}
\newtheorem{theorem}{\underline{Theorem}}[section]
\newtheorem{lemma}{\underline{Lemma}}[section]
\newtheorem{corollary}{\underline{Corollary}}[section]
\newtheorem{proposition}{\underline{Proposition}}[section]
\newtheorem{example}{\underline{Example}}[section]
\newtheorem{remark}{\underline{Remark}}[section]
\newtheorem{aspect}{\underline{Aspect}}

\begin{abstract}
Cooperative communications has long been proposed as an effective method for reducing the energy consumption of the mobile terminals (MTs) in wireless cellular networks. However, it is hard to be implemented due to the lack of incentives for the MTs to cooperate.  In this paper, we propose a pricing mechanism to incentivize  the uplink cooperative communications for the energy saving of MTs. We first consider the ideal case of MTs' full cooperation under complete information.  For this scenario as the benchmark case, where the private information of the helping MTs such as the channel and battery conditions is completely known by the source MT, the problem is formulated as a relay selection problem. Then, for the practical case of partial cooperation with incomplete information, the MTs need to cooperate under the uncertainties of the helping MTs' channel and battery conditions. For this scenario, we propose a partial cooperation scheme with pricing where a source  MT in low battery level or bad channel condition is allowed to select and pay another MT in proximity to help forward its  data to the base station (BS). We formulate  the source MT's pricing and load sharing problem as an optimization problem. Efficient algorithms based on dichotomous search and alternative optimization are proposed to solve the problem for the cases of splittable and non-splittable data at the source MT, respectively.  Finally, extensive numerical results  are provided to show that our proposed cooperative communications scheme with pricing can significantly decrease both the communications and battery outages for the MTs, and can also increase the average battery level during the MTs' operation.
\end{abstract}


\section{Introduction}
\lettrine{W}{ith} the recent developments in the smart phones and  the multimedia applications, wireless cellular network is now experiencing an exponential increase in the wireless data traffic and today's mobile terminals (MTs) consume a lot more energy than before.  Considering their limited battery capacities, MTs need to be charged more frequently and this has become the biggest customer complaint for smart phones \cite{CNN2005}. As such, reducing the energy consumption for the MT is of critical importance for resolving the energy shortage of the MTs and improving the connectivity of the wireless networks.  Furthermore, it has been shown that  the communications modules constitute a large proportion of the MTs' energy consumption, for either the MTs from the earlier 2G and 3G era \cite{Perrucci2011} or the more modern 4G mobile phones \cite{Pathak2012}.  Therefore, this gives  us a good motivation to investigate the  energy saving for the MTs in data communications.

Cooperative communications \cite{kramer2007cooperative} is an effective approach for energy saving in wireless cellular networks and wireless sensor networks. However, the battery levels and their heterogeneity among MTs/sensors  have not been rigorously considered before. For the MTs within a cellular network, some MTs are low in battery level  and others are high. If the battery level is ignored, it is possible that some MTs low in battery level still help the other MTs for data relaying. This is clearly undesirable since the battery of the MT can be easily depleted.  Under this circumstance, it would be helpful if the MTs in low battery level can get help from those high in battery level such that their operation time can be prolonged.  Hence, this motivates this work to consider  cooperative communications for energy saving  with the consideration of the battery levels of the MTs.

Furthermore, another unsolved issue in  cooperative communications is that the MTs may lack the proper incentives to cooperate. For most of the existing studies in the literature, it is assumed that the sensors in the wireless sensor networks or MTs in the wireless cellular networks cooperate with each other without self-interests. In reality, this might be true for the case of  wireless sensor networks, since the sensors within a target area usually belong to the same entity. While, this can hardly be true for the MTs in the wireless cellular networks, since the MTs belong to different individuals with self-interests. Therefore, in order to enable a practical implementation of the energy-saving cooperative communications in cellular networks, incentive design must be considered for the MTs.

\subsection{Related Work}\label{subsec:RelatedWorks}
 It is noted that there are already prior works investigating the MT-side energy saving in the literature \cite{CuiGoldsmith2005,FuKim2011,KimCeciana2010}. In particular, \cite{CuiGoldsmith2005} studied the optimal modulation scheme to  minimize the total energy consumption for transmitting a data package of a given size. Both uncoded and coded systems are considered for the modulation optimization.  \cite{FuKim2011} studies the optimal power control problem for the minimization of the average  MT energy consumption in the multi-cell TDMA system. In \cite{KimCeciana2010}, the authors study the  energy saving of the MTs by leveraging the spare capacity at the base stations (BSs) in cellular networks. The optimal design is obtained by solving the optimization problems for the scenarios of  real-time data traffic and data files transmission, respectively. Recently, \cite{LuoZhangLim2014,LuoZhangLim2015} showed that there is in general a trade-off between minimizing the energy consumption at the BSs and that at the MTs for meeting given quality of service (QoS) requirements of the MTs.

Moreover, cooperative communications for the energy saving of the MTs has been investigated in the literature of  wireless sensor or cellular networks \cite{ZhouCui2008,ZouZhuZhang2013,LiuWangGuo}. In particular, \cite{ZhouCui2008} studies the optimal timer-based relay selection scheme for the minimization of the sum energy consumption and maximization of the network lifetime.  \cite{ZouZhuZhang2013} proposed a space-time coding scheme for the MTs to cooperatively transmit to the BS under given outage and capacity requirements such that total transmit energy is minimized. {\cite{LiuWangGuo} considered the minimization of energy consumption under quality of service (QoS) constraint  with cooperative spectrum sharing in the cognitive radio network.} {\cite{ReviewReco} considered extending the lifetime of the machine-to-machine (M2M) communications network by considering the cooperative Medium Access Control (MAC) protocol.}

Although cooperative communications has long been proposed for energy saving, it is hard to be realized in reality due to the lack of incentives that motivate the MTs to cooperate \cite{YangFang2012}. {The idea of {\it virtual currency} for incentivizing the cooperation between self-organized entities was first proposed in \cite{VirtualCurrency} under the setup of wireless sensor network.} From this perspective, prior works \cite{WangHan2009,YangHuang2013,PIMRC,KandeepanJayaweera2012} have also  proposed various incentive mechanisms to motivate  cooperative communications in wireless communications systems. Specifically, \cite{WangHan2009} proposed a distributed game-theoretical framework over multiuser cooperative communication networks to achieve optimal relay selection and power allocation.  A two-stage Stackelberg game is formulated to consider the interests of the source and relay, where the source node is modeled as a buyer and the relay nodes are modeled as sellers for providing relay for the source. The difference of this work from our is that the battery level of the MTs are not considered.  \cite{YangHuang2013} studied the dynamic bargaining-based cooperative spectrum sharing between a primary user (PU) and a secondary user (SU), where the PU shares spectrum to the SU and the SU helps relay the signal of the PU in return. Different from these works, we study the energy saving of the MTs with the new consideration of the battery levels of the MTs in the cooperative communications. {\cite{PIMRC} proposed a so-called {\it reputation system} based on a reputation auction framework to provide indirect reciprocity for stimulating node cooperation in green wireless networks. The difference from our work is that it does not consider the issue of battery level in the cooperative communications and the approach for motivating the cooperation is different. \cite{KandeepanJayaweera2012} considered the business model for cooperative networking problem with the auction theory.  However, it did not give an exact modelling for the battery level and the proposed cooperative communications scheme is not in the setup of wireless cellular network.}

\subsection{Main Contributions and Organization}\label{sec:SummaryAndOrganizaiton}
The main contributions of this paper are summarized as follows:
\begin{itemize}
	\item {\it Pricing mechanism for incentivizing cooperation}: In this paper, we consider that the MTs in the network are selfish and only willing to cooperate when they can benefit from the cooperation. Different from the previous works on cooperative communications, we take the battery level of the MT into consideration and  exploit the heterogeneities of the battery levels and channel conditions between the MTs for cooperation. Under the uncertainties of the helping MTs' battery levels and channel conditions, we propose a new pricing mechanism to incentivize the cooperative communications between the MTs that can lead to a win-win situation.

	\item   {\it Full cooperation under complete information}: First, for the ideal case of full cooperation under complete information, the problem is formulated as a deterministic relay selection problem among all the helping MTs for the cases of splittable or non-splittable data at the source MT. It is further shown that in the case of splittable data, the optimal rate allocation follows a simple threshold structure and can be implemented efficiently.

	\item {\it Partial cooperation under incomplete information}: Then, for the practical case of partial cooperation under incomplete information, the MTs belong to entities of individual interests and cannot share private information to the other MTs. Under the uncertainties on the battery levels and channel conditions of the helping MTs,  we formulate the MT's pricing and load sharing problem as an optimization problem for the two cases of splittable and non-splittable data of the source MT, respectively. Efficient algorithms based on dichotomous search and alternative optimization are proposed for the solutions of the problem.

\end{itemize}

 The rest of this paper is organized as follows. Section \ref{sec:system_model} introduces the system model with the  cooperative communications for MTs' energy saving, and the resulting cost and utility functions.  Section \ref{sec:completecoop} discusses our proposed protocol under complete information as the performance benchmark and  Section \ref{sec:OptimalP2} studies the general case of  cooperative communications under incomplete information. Section \ref{sec:numerical_examples} presents numerical examples to validate the results in this paper. Finally, Section \ref{sec:conclusion} concludes this paper  and discusses future work.

\section{System Model and Energy-saving Cooperation}\label{sec:system_model}

\begin{figure}[t]
  \centering
  \includegraphics[width=9cm]{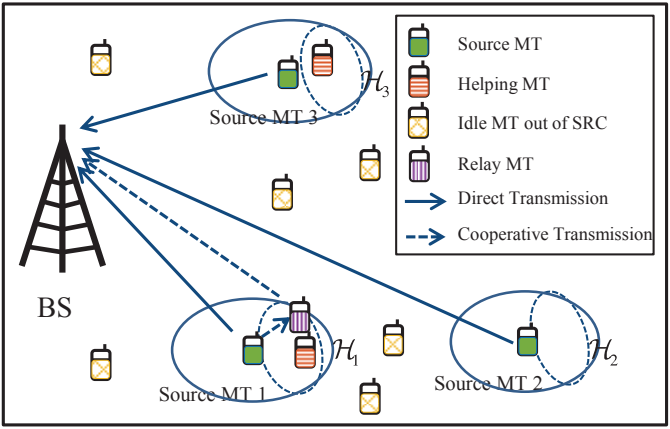}
  \caption{System model for direct and cooperative data transmission.}\label{system_model}
\end{figure}
\begin{table}[htp]\caption{List of notations and their physical meanings.}
\centering
\begin{tabular}{l  L{7.2cm}@{} }
\toprule
Symbols&Physical Meanings\\
\cmidrule(l){1-1}\cmidrule(l){2-2}
$\mathcal{K}$ & Set of all the MTs \\
$\mathcal{K}_S$  &  Set of the source MTs \\
$\mathcal{K}_I$  &   Set of the idle MTs\\
$\mathcal{H}_i$  & Set of helping MTs for source MT $i\in\mathcal{K}_S$\\
$\rho$& Probability that a certain MT initiates data traffic\\
$\mu_{N_i}$ & Average number of helping MTs for source MT $i\inm{K}_S$\\
$h_k$ & Channel coefficient of MT $k\inm{K}$\\
$g_k$& Channel gain of MT $k\inm{K}$\\
$r_0$& Reference distance\\
$r_k$& Distance from MT $k\inm{K}$ to the BS\\
$\alpha$& Exponent of the large-scale power attenuation \\
$G_0$& Pathloss at reference distance $r_0$\\
$G_k$&  Pathloss for MT $k\inm{K}$ \\
$D_i$ & Data rate of source MT $i\inm{K}_S$\\
$D_i^{(S)}$& Data rate of the source MT $i\inm{K}_S$ in the CT mode\\
$D_i^{(R)}$ &Data rate of the relay MT $j\inm{H}_i$ in the CT mode\\
$\sigma^2$& Power of the noise at the receiver of the BS\\
$B_k$ & Battery level of MT $k\in\mathcal{K}$\\
${E}_i^{(D,S)}$& Energy consumption of source MT $i\inm{K}_S$ with DT mode\\
$E_i^{(C,S)}$& Energy consumption of  source MT $i\inm{K}_S$ with CT mode\\
 $E_j^{(C,R)}$& Energy consumption of  helping MT $j\inm{H}_i$ with CT mode\\
$\zeta_k$  &Unit energy cost  for MT $k\in\mathcal{K}$\\
$\pi_i$ & Payment from the source MT $i\inm{K}_S$ to its helping MTs\\
$U_{j}$  & Utility for the helping MT $j\inm{H}_i$\\
$C_i$ & Cost for the source MT $i\inm{K}_S$  with CT  mode\\
$\epsilon$& Utility margin for the relay MT\\
$\eta_k$ &Exponential distributed Rayleigh fading power for MT $k\inm{K}$\\
$\gamma_i$& Cost reduction threshold of the source MT $i\inm{K}_S$\\
$C_{i,j}$& Source MT $i$'s cost associating  with helping MT $j\inm{H}_i$\\
\bottomrule
\end{tabular}
\label{tab:TableOfNotationForMyResearch}

\end{table}
\subsection{System Model}
All the notations used in this paper are summarized and explained in Table \ref{tab:TableOfNotationForMyResearch} for the ease of reading. As shown in Fig. \ref{system_model}, we consider the uplink data transmission within one single cell of a cellular network.\footnote{Our results can be extended to the case of multiple cells  by applying our results to each cell independently.} Different roles of the MTs will be introduced later in the paper. Within the cell, there is one single-antenna BS serving  $K$ single-antenna MTs denoted by the set $\mathcal{K}=\{1,2,\cdots,K\}$.  We assume that the locations of the MTs follow a two-dimensional Homogeneous Poisson Point Process (HPPP) with spatial density $\lambda$.\footnote{For the spatial user density $\lambda$, it can be readily obtained by dividing the total number of MTs within the cell over the total area of the cell. The number of the MTs can be estimated by the history data of the cell or by real-time monitoring.} \cite{StochasticGeometry}  We consider that the MTs within the cell initiate their data traffic independently with probability $\rho$. Then, according to the Marking Theorem \cite{Kingman1993}, these {\it source} MTs (i.e. MTs initiating data traffic) also form an HPPP  with density $\rho\lambda$ and the remaining {\it idle} MTs form another HPPP with density $(1-\rho)\lambda$. We denote these sets of  source MTs and idle MTs as $\mathcal{K}_S$ and $\mathcal{K}_I$, respectively, such that $\mathcal{K}_S\cup\mathcal{K}_I=\mathcal{K}$ and $\mathcal{K}_S\cap\mathcal{K}_I=\emptyset$.

We consider the uplink data transmission of all MTs and assume the narrow-band block fading channel model. To support multiple MTs, orthogonal data transmission is assumed, e.g., by applying orthogonal frequency-division multiple access (OFDMA). We denote  the complex baseband channel coefficient from MT $k\inm{K}$ to the BS as $h_k$, which follows a simplified channel model incorporating the large-scale power attenuation with loss exponent $\alpha>2$ and the small-scale  Rayleigh fading.  More specifically, we denote $r_k$ as the distance between  MT $k\inm{K}$ and the BS, and $r_0$ as a reference distance, respectively. Then, the channel coefficient $h_k$ is expressed as
\begin{align}\label{sys_model:CH}
  h_k=
  \begin{cases}
   \bar h_k \sqrt{G_0\left(\frac{r_k}{r_0}\right)^{-\alpha}},&r_k>r_0\\
	\bar h_k \sqrt{G_0},&\mathrm{otherwise}
   \end{cases}
  ,~k\in\mathcal{K},
\end{align}
where $\bar h_k\sim \mathcal{CN}(0,1),~k\inm{K}$ is an independent and identically distributed (i.i.d.) circularly symmetric complex Gaussian (CSCG) random variable with zero mean and unit variance modeling  the small-scale Rayleigh fading,  and  $G_0$ is  the constant path-loss between the MT and the BS at the reference distance $r_0$. Therefore, the channel power gain between the MT $k$ and the BS is
\begin{align}\label{channelmodel}
g_k=|h_k|^2=\eta_kG_k,~k\inm{K}.
\end{align}
Here, we denote $\eta_k\sim \mathrm{exp}(1)$ as an exponential random variable with unit mean modeling the power envelope of the Rayleigh fading and
\begin{align}
G_k=\begin{cases}
G_0\left(\frac{r_k}{r_0}\right)^{-\alpha},&r_k>r_0\\
G_0,&\mathrm{otherwise}
\end{cases}
,~k\inm{K}
\end{align}
as the power attenuation between the BS and the MT $k$ at the distance of $r_k$.

{For simplicity, we consider a time-slotted system, where symbols for the message are transmitted in each time slot. For convenience, the number of symbols transmitted per time slot are normalized to unity.} If MT $k\inm{K}$ initiates its data traffic, a message from the set $\{1,2,\cdots,2^{D_k}\}$ is sent, where  $D_k$ is the transmitted rate in bits per symbol. Without loss of generality, we also normalize the duration of one symbol time to unity such that the two terms energy and power can be used inter-changeably in the paper.  Then, if the achievable data rate $D_k$ is normalized by the available bandwidth at the MT, for given transmission energy per symbol $E_k$, the (normalized) achievable data rate for MT $k\inm{K}$ in bits/sec/Hz (bps/Hz) is
\begin{align}\label{transmissionrate}
D_k=\log_2\left(1+\frac{g_kE_k}{\sigma^2}\right),
\end{align}
where $\sigma^2$ denotes the power of the noise at the receiver of BS.

For source MT $i\inm{K}_S$, in order to accomplish the uplink transmission at (normalized) data rate $D_i$, it can choose between the following two transmission modes.
\subsubsection{Direct Transmission Mode (DT Mode)}~\\

\vspace{-1em}

In this  mode,  the source MT transmits to the BS directly with normalized data rate $D_i$.  Hence, according to (\ref{transmissionrate}) the required energy per symbol for transmitting with data rate $D_i$ is
\begin{align}\label{directTrans}
 {E}_i^{(D,S)}=\frac{\sigma^2}{g_i}\left(2^{ D_i}-1\right),~i\in\mathcal{K}_S.
\end{align}

\subsubsection{Cooperative Transmission Mode (CT Mode)}~\\

\vspace{-1em}

In this mode, for a certain source MT $i\in\mathcal{K}_S$, as shown in Fig. \ref{system_model}, it can {\it associate} with one idle MT (if any) within the distance  $d$ as its {\it relay} MT that can help relay the data to the BS, where $d$ is the range of the short range communications (SRC) such as WiFi-Direct\cite{WifiDirect}, Bluetooth\cite{bluetooth}, etc.\footnote{In this paper, we assume single relay selection to keep the overhead low. Similar approach has been used in \cite{ZhouCui2008}.} We denote this set of idle MTs within the distance $d$ from the source MT $i\inm{K}_S$ as its set of {\it helping} MTs $\mathcal{H}_i\subset\mathcal{K}_I$, where $|\mathcal{H}_i|=N_i$ is the number of MTs within the set.  Then, it follows that $N_i$ is a Poisson random variable with mean $\mu_{N_i} =(1-\rho)\lambda \pi d^2,~i\inm{K}_S$ and its probability mass function (PMF) is given by
\begin{align}\label{HPPP}
  \mathrm{Pr}(N_i=n)=\frac{\mu_{N_i}^n}{n!}e^{-\mu_{N_i}},~n=0,1,\cdots,~i\in\mathcal{K}_S.
\end{align}
From (\ref{HPPP}), we observe that the PMF of $N_i$ is proportional to the range of the SRC $d$, an MT's probability of remaining idle $1-\rho$ and the spatial density $\lambda$. Note that if $N_i=0$ or $\mathcal{H}_i=\emptyset$, source MT $i\inm{K}_S$ will operate in DT mode, i.e., transmit directly to the BS; while if $N_i\geq 1$, source MT $i\inm{K}_S$ can operate in CT mode by selecting one from its helping MTs in $\mathcal{H}_i$ to relay the data.

For the CT mode,  the source MT $i\inm{K}_S$  in general splits data $ D_i$ into two parts with $D_i=D_i^{(S)}+D_i^{(R)}$: $D_i^{(S)}$ for the source MT to transmit directly to the BS and $D_i^{(R)}$ for its relay MT to transmit. For transmitting the data $D_i^{(S)}$, similar to (\ref{directTrans}), the required energy for the source MT $i\inm{K}_S$ is\footnote{In this paper, we do not directly consider the maximum power constraint of the MT in order to obtain tractable problem formulation and insights. However, as will be shown later, we have already implicitly considered the issue of large transmit power. When the transmit power is very large, it incurs a large cost on the MT and the MT will try to get help from the other MTs. Hence, large energy consumption can be avoided. The simulation results in Section \ref{SystemSimulation} will corroborate the effectiveness of our scheme.}

\begin{align}
{E}_i^{(C,S)}=\frac{\sigma^2}{g_i}\left(2^{ D_i^{(S)}}-1\right),~i\in\mathcal{K}_S.
\end{align}
Then, for the other part of data $D_i^{(R)}$, as shown by the dashed line in Fig. \ref{system_model}, the source MT first transmits it to the selected relay MT and then the relay MT decodes and forwards the signal to the BS. In practice, SRC technologies (e.g. WiFi-Direct\cite{WifiDirect}, Bluetooth\cite{bluetooth}, etc.) offer high communications data rate with low transmit power. The energy consumption and the transmission time is also small compared to that in the wireless cellular network. Hence, we ignore them for this short range data transmission.\footnote{For example, the maximum transmission power of WiFi-Direct is $30~\mathrm{mW}$ and the data rate can be as high as $250~\mathrm{Mbps}$ \cite{WifiDirect}. While for the LTE mobile terminal in wireless cellular network, the typical transmit power is $200~\mathrm{mW}$ and the peak data rate is $75~\mathrm{Mbps}$\cite{LTEstandard}. Hence, the transmit power or the duration of the SRC between MTs is much lower compared to that of the cellular communications in the uplink and can be ignored. Furthermore, the analysis can be easily  extended to the case that the energy consumption  and transmission time of SRC are constants and there will not be major changes in the results.}  Also due to the small range, the source MT $i\inm{K}_S$ and its helping MT $j\in \mathcal{H}_i$ have roughly the same distance to the BS (i.e. $r_i=r_j$) and can be assumed to have the same path-loss. Hence, the channel power gain between the helping MT $j\inm{H}_i$ of the source MT $i\inm{K}_S$ and the BS is
\begin{align}\label{ChannelGain}
g_j=\eta_jG_i,
\end{align}
where the short-term Rayleigh fading of the channel power $\eta_j$ is still independently distributed among the MTs.  Hence, if helping MT $j\inm{H}_i$ is selected as the relay MT, the energy consumption for this data transmission is
\begin{align}\label{energy_Consume_ct}
 E_j^{(C,R)}=\frac{\sigma^2}{g_j}\left(2^{ D_i^{(R)}}-1\right),~j\in\mathcal{H}_i, ~i\in\mathcal{K}_S.
\end{align}

 \subsection{Definition of  Costs and Utilities}\label{sec:problem_formulation}
 At different battery levels, an MT has different valuations of the remaining energy in its battery.  The energy stored in the battery is generally more valuable when the battery level is low. Hence, we define the unit energy cost $\zeta_k$ for each MT $k\inm{K}$ as a function of its battery level $B_k$, i.e.,
\begin{align}
\zeta_k=f(B_k),\label{price_function}
\end{align}
where $B_k\in[0,B_{\mathrm{max}}]$ is the battery level of MT $k$ with its range from zero to the maximum storage $B_{\mathrm{max}}$\footnote{For analytical tractability, in this paper, we assume that all the MTs have the same battery capacity $B_{\mathrm{max}}$. }, and $f:[0,B_{\mathrm{max}}]\rightarrow [0,\zeta_{\mathrm{max}}]$ is a monotonically decreasing function of $B_k$ whose range is from zero to the maximum energy cost $\zeta_{\mathrm{max}}>0$.\footnote{This design of function $f$ is reasonable as a user will value energy more when facing low battery, and we assume the minimum energy cost equal to zero when $B_k=B_{\mathrm{max}}$.}

In order to motivate the helping MT's participation in the cooperation, if a helping MT $j\in\mathcal{H}_i$ is selected by the source MT $i\inm{K}_S$ as the relay MT, it will receive a price $\pi_i$ for transmitting  with data rate $D_i^{(R)}$. The payment can be in the form of currency or credits in a multimedia application. Hence, the utility of helping MT $j\inm{H}_i$ by participating in the cooperation is $\pi_{i}-\zeta_j\ejr$, where $\ejr$ is the energy consumption for transmitting with data rate $D_i^{(R)}$ as defined in (\ref{energy_Consume_ct}). Furthermore, the helping MT has a  reservation utility of $\epsilon\geq 0$ for accepting the request. That is, helping MT $j\inm{H}_i$ will only accept the relay request from source MT $i\inm{K}_S$ if $\pi_{i}-\zeta_j\ejr\geq \epsilon$. Therefore, the utility of the helping MT $j\in\mathcal{H}_i$ for source MT $i\inm{K}_S$ is the difference between the price and the energy cost if the difference is larger than $\epsilon$ and zero otherwise, which is defined as\footnote{Here, note that the utility function is a concave function of $D_i^{(S)}$ with diminishing return and the cost function to be defined in (\ref{formulaiton_source_utility}) is a convex and monotonically increasing function with respect to $D_i^{(S)}$. Hence, these definitions conform to the classic definition of cost and utility functions in economics \cite{mas1995}.}
\begin{align}\label{formulaiton_idling_utility}
U_{j}=
\begin{cases}
\pi_{i}-\zeta_j\ejr,&\mathrm{if}~\pi_{i}-\zeta_j\ejr\geq \epsilon,\\
0,&\mathrm{otherwise.}
\end{cases}.
\end{align}
For the source MT $i\inm{K}_S$, if there is at least one helping  MT  accepting the price $\pi_i$, the cost of the source MT $i\inm{K}_S$  is the sum of the price $\pi_i$ and the energy cost by direct transmission $\zeta_i\eis$. Otherwise, it needs to directly transmit to the BS with rate $D_i$   at the cost of $\zeta_iE_i^{(D,S)}$. Thus, the energy cost of source MT $i\inm{K}_S$ is
\begin{align}\label{formulaiton_source_utility}
 C_{i}=
\begin{cases}
\pi_{i}+\zeta_i\eis ,&\mathrm{if}~ \exists j\in\mathcal{H}_i, \pi_{i}-\zeta_jE_j^{(C,R)} \geq \epsilon, \\
\zeta_iE_i^{(D,S)},& \mathrm{otherwise.}
\end{cases}.
\end{align}
To ensure the mutual benefits of the source MT $i\inm{K}_S$ and relay MT $j\inm{H}_i$ in the cooperation, the price $\pi_i$ should  satisfy the following inequality
\begin{align}\label{CoopInequality}
\epsilon \overset{(a)}{\leq}\pi_i\overset{(b)}{\leq}  \zeta_iE_i^{(D,S)}-\zeta_i\eis,
  \end{align}
where inequality (a) ensures the utility increase of the helping MTs in the cooperation  and inequality (b) ensures cost reduction for the source MT. Note that $\zeta_iE_i^{(D,S)}\geq \epsilon$ must hold  for the feasibility of the CT mode. That is, the value of the energy consumption by direct transmission at the source MT must be larger than the reservation utility of the helping MT. 

\subsection{Cooperative Transmission Protocol}\label{subsec:ModeSelection}
\begin{figure}[t]
  \centering
  \includegraphics[width=7cm]{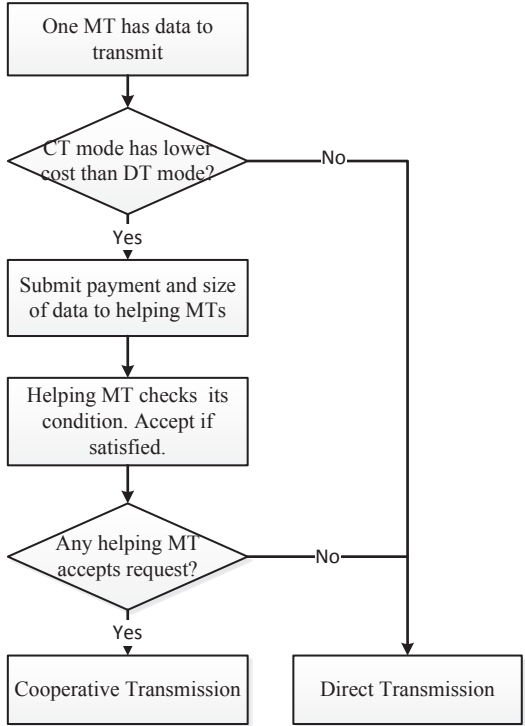}
  \caption{Cooperative communications protocol.}\label{comcoop}
\end{figure}
Next, in order for the MTs in the cellular network to cooperate with mutual benefits, we propose the following  cooperative data transmission protocol, which is also shown by the flow chart in Fig. \ref{comcoop}.

\begin{itemize}
  \item[1)] When an MT has data to transmit, it chooses between the CT mode and DT mode according to the criterion to be specified later {in (\ref{compinfocondition}) and (\ref{condition:CTmode}), for the cases of complete and incomplete information, respectively.}
  \item[2)]  If the DT mode is selected, the source MT transmits directly to the BS. If the CT mode is selected, it broadcasts the proposed payment and the relay data rate to all its helping MTs.
  \item[3)] The helping MT (if any)  accepts the request and sends an acceptance notification to the source MT if the condition for cooperation is satisfied or rejects the request otherwise.
  \item[4)]  If multiple helping MTs accept the relay request,  the source MT randomly chooses one MT as the relay MT and  transmits the data with the CT mode.\footnote{Because the relay data rate and price are already determined by the source MT and each relay candidate provides the same help to the source MT, the source MT does not care about which helping MT among those accepting the request is chosen.} Otherwise, the source MT transmits with the DT mode.
\end{itemize}

In the above proposed cooperative transmission protocol, the key challenge is the mechanism design for incentivizing the cooperation of the MTs such that the MTs can mutually benefit. In the following sections, we propose pricing-based incentive mechanism design for the cooperation under different information sharing scenarios.

 \section{Benchmark Case: Full Cooperation under Complete Information}\label{sec:completecoop}
In this section, we consider the ideal case of full cooperation under complete information, where the private information of the helping MTs $j\inm{H}_i$, including the number of helping MTs $N_i$, their battery levels $B_j$'s and channel conditions $g_j$'s, is known by each source MT $i\inm{K}_S$. This case can happen when the MTs belong to a fully cooperative group (e.g., friends) that they are willing to help each other without the requirement on the reservation utility $\epsilon$ and share their private information truthfully. This case will also provide the performance benchmark (upper bound) for the  partial cooperation under incomplete information in the next section.

Due to the full cooperation nature among the MTs, the reservation utility of the helping MT $\epsilon$ reduces to zero. Hence, source MT $i\inm{K}_S$ only needs to give a payment to its helping MT $j\inm{H}_i$ that is just enough to cover the cost $\zeta_jE_j^{(C,R)}$ for transmitting $D_i^{(R)}$ such that the helping MT's utility in (\ref{formulaiton_idling_utility}) is non-negative. Hence, the required amount of payment to helping MT $j\inm{H}_i$ from source MT $i\inm{K}_S$ is $\zeta_jE_j^{(C,R)}$. Then, source MT $i\inm{K}_S$ needs to optimize the relay data rate $D_i^{(R)}$ for each helping MT $j\inm{H}_i$ to minimize the sum energy cost, i.e.,
\begin{align}\label{CompInfoCost}
C_{i,j}=\mathop{\mathtt{min.}}_{D_i^{(R)}\geq 0}&~\zeta_jE_j^{(C,R)}+\zeta_iE_i^{(C,S)}\nonumber\\
\mathrm{s.t.}~&~D_i^{(R)}+D_i^{(S)}=D_i.
\end{align}
Problem (\ref{CompInfoCost}) can be considered as a weighted sum energy minimization problem for the source and helping MTs, where the weight is the unit energy cost of the individual MT. It is evident that when the weights (i.e. unit energy cost) of the source and helping MTs are equal, this problem reduces to the sum energy minimization problem. When the weight  of one MT is larger than the other, the problem is more favorable for the MT with lower energy and the optimization is more similar to the max-min optimization of the battery levels.

After obtaining the minimum sum cost $C_{i,j}$ of associating with each helping MT $j\inm{H}_i$, source MT $i\inm{K}_S$ chooses the best helping MT with the following relay selection problem:
\begin{align}
\mathrm{(P1)}:~\hat C_{i}=\mathop{\mathtt{min.}}_{j}&~C_{i,j},
\end{align}
where $C_{i,j}$ is obtained in problem (\ref{CompInfoCost}).

Next, we discuss the criterion for the mode selection of source MT $i\inm{K}_S$, which has been introduced in Section \ref{subsec:ModeSelection}.  For the source MT to choose the CT mode, its cost reduction from the  direct transmission  must be larger than a threshold denoted by $\gamma_i$, which accounts for the overheads in cooperative communications such as signaling and signal processing. Hence, if source MT $i\inm{K}_S$ chooses the CT mode, the following condition has to be satisfied:
\begin{align}\label{compinfocondition}
\hat C_{i}\geq \zeta_iE_i^{(D,S)}+\gamma_i.
\end{align}
In the following, we discuss the solution for the minimum cost $\hat C_{i}$ of full cooperation under complete information in two cases: non-splittable data (i.e. $D_i^{(R)}=D_i$) and splittable data (i.e. $0\leq D_i^{(R)}\leq D_i$).

\subsection{Cooperation with Non-splittable Data}\label{subsec:Incompnsd}
First, we discuss the case where the data is not splittable at the source MT due to reasons such as lack of necessary processing functionalities. In this case, all the data of the  cooperative communications is transmitted by the relay MT (i.e., $\ris=0$ and $\rir=D_i$). Hence, according to (\ref{CompInfoCost}), the cost of the source MT $i\inm{K}_S$ by associating with helping MT $j\inm{H}_i$ reduces to $C_{i,j}=\zeta_j\frac{\sigma^2}{g_j}\left(2^{D_i}-1\right)$ for problem $\mathrm{(P1)}$. Then, the minimum cost of full cooperation with non-splittable data can be obtained by solving the simplified relay selection problem in $\mathrm{(P1)}$.

Therefore, the optimal transmission of the source MT in the case of non-splittable data follows a two-step procedure: First, the source MT computes and finds the helping MT (if any) with the least energy cost. Then, it checks the condition in (\ref{compinfocondition}) and chooses between the DT mode and CT mode.

\subsection{Cooperation with Splittable Data}\label{subsec:compinfosd}
It can be proved that problem (\ref{CompInfoCost}) is a convex optimization problem and  the optimal solution is given by the following proposition.
\begin{proposition}\label{Prop:CompInfoNS}
The optimal data rate transmitted by the relay MT in problem (\ref{CompInfoCost}) is given by
\begin{align}
{\hat D_i}^{(R)}=
\begin{cases}
0,& \mathrm{if}~ \log_2\frac{\theta_i}{\theta_j}<-D_i\\
\frac{1}{2}(D_i+\log_2\frac{\theta_i}{\theta_j}),& \mathrm{if}~ -D_i \leq \log_2\frac{\theta_i}{\theta_j}< D_i\\
D_i,& \mathrm{if}~ \log_2\frac{\theta_i}{\theta_j}\geq D_i\\
\end{cases},
\end{align}
where $\theta_i=\frac{\zeta_i}{\eta_i}$ and $\theta_j=\frac{\zeta_j}{\eta_j}$ can be interpreted as the {\it effective energy cost} of the source MT $i\inm{K}_S$ and helping MT $j\inm{H}_i$, respectively.
\end{proposition}
\begin{proof}
Please refer to Appendix \ref{appen:prop} for the details.
\end{proof}

{It can be observed from Proposition \ref{Prop:CompInfoNS} that the optimal relay data rate follows a threshold structure with respect to the log-ratio between the effective energy costs of the source and helping MTs. When the effective energy cost $\theta_i$ of the source MT $i\inm{K}_S$ is much lower than that of the relay MT $j\inm{H}_i$ to the extent that $\log_2\frac{\theta_i}{\theta_j}<-D_i$ is satisfied, the source MT will not ask for help from this helping MT and transmit all by itself. If the effective energy cost of the source and helping MT is comparable, then the source and helping MT will split the data package $D_i$ for transmission. Finally, if the effective energy cost of the source is much higher than that of the helping MT so that $\log_2\frac{\theta_i}{\theta_j}\geq D_i$ is satisfied, then the helping MT will transmit the whole data package.
}

\section{General Case: Partial Cooperation under Incomplete Information}\label{sec:OptimalP2}

In the previous section, we have considered the full cooperation under complete information, which is the optimal scenario for the source MT and can serve as the benchmark scheme. However, this scenario is not applicable if the MTs belong to different entities that are not fully cooperative and are unwilling to share private information to each other. In this section, we consider the general scenario where the MTs do not know exactly the other MTs' channel condition and battery level and discuss how these MTs still can cooperate with mutual benefits under this scenario.

\subsection{Problem Formulation}
For cooperation under incomplete information between the source and helping MTs, we formulate the problem of decision making under uncertainties with the expected utility theory \cite{fishburn1970utility}.  We denote $\mathrm{Pr}(\pi_{i}-\zeta_j E_j^{(C,R)} \leq \epsilon)$ as the probability that helping MT $j\inm{H}_i$ rejects the request given by the source MT $i\inm{K}_S$. We assume that all the channel gains $g_j$'s and battery states $B_j$'s of the helping MTs $j\inm{H}_i$ are independent.  Hence, given the set of helping MTs $\mathcal{H}_i$, the conditional expected cost of the source MT $i\inm{K}_S$ for transmitting at data rate $D_i$  is
\begin{align}\label{condcost}
\small
&\mathbb{E}[C_{i} \vert \mathcal{H}_i]=\left(1-\prod_{j\inm{H}_i} \mathrm{Pr}(\pi_{i}-\zeta_j E_j^{(C,R)} \leq \epsilon)\right)\nonumber\\
&\times (\pi_i+\zeta_i\eis)+\left(\prod_{j\inm{H}_i} \mathrm{Pr}(\pi_{i}-\zeta_j E_j^{(C,R)} \leq \epsilon)\right)\zeta_iE_i^{(D,S)}\nonumber\\
&=\left(1- \prod_{j\inm{H}_i} \mathrm{Pr}(\pi_{i}-\zeta_j E_j^{(C,R)} \leq \epsilon)\right)\nonumber\\
&\times(\pi_i+\zeta_i\eis-\zeta_iE_i^{(D,S)})
+\zeta_iE_i^{(D,S)},
\end{align}
where the expectation is taken over the two possible outcomes of successful and unsuccessful relay association in (\ref{formulaiton_source_utility}). By further considering all possibilities of helping MT set $\mathcal{H}_i$ for source MT $i\inm{K}_S$ in (\ref{HPPP}),  the expected cost of the source MT $i\inm{K}_S$ can be obtained by applying the law of iterated expectation, i.e.,
\begin{align}\label{expectedcost}
\mathbb{E}[C_i]&=\mathbb{E}[\mathbb{E}[C_{i}\vert \mathcal{H}_i]]\nonumber\\
&=\sum_{n=0}^{\infty} \mathrm{Pr}(N_i=n)\mathbb{E}[C_{i}\vert \mathcal{H}_i],~i\inm{K}_S,
\end{align}
{Here, it is worthwhile to discuss the role of reservation utility $\epsilon$ in the expected energy cost $\mathbb{E}[C_i]$. As $\epsilon$ denotes the level of minimum benefit for the relay MT in the cooperation, it can be observed that the expected energy cost $\mE[C_i]$ should be monotonically increasing with $\epsilon$. That is, with a higher reservation utility for the relay MT, the expected energy cost of the source MT is also higher.}

Then, we formulate the optimization problem that minimizes the expected cost of the source MT $i\inm{K}_S$ over the price $\pi_i$ and relay data $\rir$ as follows:
\begin{align}
\mathrm{(P2)}:~&\mathop{\mathtt{min.}}_{\pi_{i},\rir\geq 0}~~ \mathbb{E}[C_{i}]\nonumber\\
\mathtt{s.t.}
&~~ \epsilon\leq\pi_i\leq  \zeta_iE_i^{(D,S)}-\zeta_i\eis,\label{PaymentConstraint}\\
&~~ \ris+\rir= D_i.
\end{align}

Next, we discuss the criterion for the mode selection between the DT mode and CT mode. Similar to the condition for the full cooperation case in (\ref{compinfocondition}), for choosing the CT mode, we require the (expected) reduction of the source MT's energy cost from that of the direct transmission to be larger than a threshold $\gamma_i$. In addition, considering the feasibility condition for cooperation in (\ref{CoopInequality}),  the condition for the source MT to choose the CT mode  is
 \begin{align}\label{condition:CTmode}
 \zeta_iE_i^{(D,S)}\geq \max\{\gamma_i+\mathbb{E}[{C_i^*}],\epsilon\},
 \end{align}
where  $\mathbb{E}[{C_i^*}]$ is the minimum expected cost obtained in problem $\mathrm{(P2)}$. It should be noted that the problem $\mathrm{(P2)}$ is hard to be proved to be convex due to its complex objective function; thus, it is difficult to obtain its optimal solution in general.  In the following two subsections, similar to the previous section, we discuss the minimum expected cost $\mathbb{E}[{C_i^*}]$ of the source MT $i\inm{K}_S$ in details depending on whether the data is splittable or not. 

\subsection{Proposed Solution for Problem (P2)}
In this subsection, we first simplify problem $\mathrm{(P2)}$ under some further assumptions. Then, similar to Section \ref{sec:completecoop}, we discuss the solution of the problem under the cases of non-splittable and splittable data, respectively. With the energy consumption $E_j^{(C,R)}$ for the helping MT $j\inm{H}_i$ defined in (\ref{energy_Consume_ct}), the probability of successful association between source MT $i\inm{K}_S$ and its helping MT $j\in\mathcal{H}_i$ in (\ref{condcost}) is
\begin{align}
\mathrm{Pr}(\pi_i-\zeta_j E_j^{(C,R)} \geq \epsilon)&=\mathrm{Pr}\left( \frac{\zeta_j}{\eta_j} \leq \frac{G_i(\pi_i-\epsilon)}{\sigma^2(2^{\rir}-1)}\right)\nonumber\\
&=\mathrm{Pr}(\frac{\zeta_j}{\eta_j}\leq w_i),\label{theta_CDF}
\end{align}
where $w_i$ is denoted as
\begin{align}\label{importantparam}
 w_i=\frac{G_i(\pi_i-\epsilon)}{\sigma^2(2^{\rir}-1)}.
 \end{align}

 For simplicity, we further assume that the relation between an MT's unit energy cost $\zeta_k$ and its battery level $B_k$ in (\ref{price_function}) follows a linear function\footnote{It should be noted that our analysis can be extended to the other monotonically non-increasing functions, whose analysis will be technically more involved but offers essentially similar engineering insights. It should also be noted that the choice of the function $f$  reflects the sensitivity of the MTs towards the usage of the energy in the battery. By adopting a function that is {\it in-different} to the battery level, the design objective is more similar to minimizing the total energy consumption. Instead, by adopting a function that is {\it sensitive} to the battery level, this design is more favorable for the MTs with low battery level.}
\begin{align}\label{linearfunction}
\zeta_k=\zeta_{\mathrm{max}}\left(1-\frac{B_k}{B_{\mathrm{max}}}\right).
\end{align}
 We also assume that the battery level $B_j$ of the helping MT $j\inm{H}_i$ is known to the source MT $i\inm{K}_S$ as uniform distribution, i.e. $B_j\sim\mathcal{U}[0,B_\mathrm{max}]$.\footnote{Our proposed scheme can still be applicable to the case of heterogeneous battery capacity. One heuristic is that, based on the statistics of the battery capacities of the MTs, the source MT $i\inm{K}_S$ can obtain the average battery capacities of the MTs as $\bar{B}_{\mathrm {max}}$ and predict the battery level of the helping MT $j\inm{H}_i$ as $B_j\sim\mathcal{U}[0,\bar{B}_{\mathrm{max}}]$. Then, the proposed cooperative communications protocol with pricing under uncertainty still applies.} Then, due to the linear function in (\ref{linearfunction}), the energy cost $\zeta_j$ is also uniformly distributed as $\zeta_j\sim\mathcal{U}[0,\zeta_{\mathrm{max}}]$. Hence, the probability of successful association between the source MT $i\inm{K}_S$ and helping MT $j\inm{H}_i$  is
\begin{align}\label{SuccessfulAssociation}
&\mathrm{Pr}(\pi_i-\zeta_j E_j^{(C,R)} \geq \epsilon)=\mathrm{Pr}(\eta_j\geq \frac{\zeta_j}{w_i})\nonumber\\
&=\frac{1}{\zeta_{\mathrm{max}}}\int_{0}^{\zeta_{\mathrm{max}}}\int_{\frac{\zeta_j}{w_i}}^{\infty} e^{-\eta_j}d \eta_jd\zeta_j=\frac{w_i}{\zeta_{\mathrm{max}}}(1-e^{-\frac{\zeta_{\mathrm{max}}}{w_i}}).
\end{align}
With the results in (\ref{condcost}) and (\ref{SuccessfulAssociation}), the objective of problem {$\mathrm{(P2)}$} in (\ref{expectedcost}) can be simplified as
\begin{align}\label{newobj}
\small
\mathbb{E}[C_i]&=\sum_{n=0}^{\infty}\mathrm{Pr}(N_i=n)\Bigg\{\left[1-\left(1-\frac{w_i}{\zeta_{\mathrm{max}}}\left(1-e^{-\frac{\zeta_{\mathrm{max}}}{w_i}}\right)\right)^{n}\right]\nonumber\\
&\times \left(\pi_i+\zeta_i\eis- \zeta_iE_i^{(D,S)}\right)+ \zeta_iE_i^{(D,S)}\Bigg\}.
\end{align}

Next, we discuss the convexity of problem {$\mathrm{(P2)}$}  by the following proposition.
{ \begin{proposition}\label{proposition:concave}
Problem {$\mathrm{(P2)}$} is marginally convex with respect to $\pi_i$ and $\rir$.
\end{proposition}}
\begin{proof}
Please refer to Appendix \ref{Sec:Appen} for the details.
\end{proof}

{It should be noted that problem $\mathrm{(P2)}$ is not a convex optimization problem. This is  because the objective of the problem is not  jointly convex with respect to $\pi_i$ and $D_i^{(R)}$.} In the following, similar to Section \ref{sec:completecoop}, we discuss the optimal solution for problem {$\mathrm{(P2)}$} to obtain $\mathbb{E}[{C_i^*}]$ under two cases: non-splittable (i.e. $D_i^{(R)}=D_i$) and splittable data (i.e. $0\leq D_i^{(R)}\leq D_i$).
\vspace{0.5em}
\subsubsection{Optimal Pricing for Non-splittable Data}\label{subsec:woOLS}~\\

\vspace{-1em}
First, we discuss the case where the data is not splittable. In this case, all the data of the source MTs is transmitted by the relay MT (i.e., $\ris=0$ and $\rir=D_i$). As $w_i$ in (\ref{importantparam}) is now reduced to $w_i=\frac{G_i(\pi_i-\epsilon)}{\sigma^2(2^{D_i}-1)}$, problem {$\mathrm{(P2)}$} is simplified to the following problem without load sharing:
\begin{align}
&\mathrm{(P2')}:\nonumber\\
\mathop{\mathtt{min.}}_{\pi_i}&~\sum_{n=0}^{\infty}\mathrm{Pr}(N_i=n)\Bigg\{\left[1-\left(1-\frac{w_i}{\zeta_{\mathrm{max}}}\left(1-e^{-\frac{\zeta_{\mathrm{max}}}{ w_i}}\right)\right)^{n}\right]\nonumber\\
&\times\left(\pi_i-\zeta_iE_i^{(D,S)}\right)+\zeta_iE_i^{(D,S)}\Bigg\} \nonumber\\
 \mathtt{s.t.}
 &~\epsilon\leq \pi_i\leq  \zeta_iE_i^{(D,S)}\nonumber.
\end{align}
Because the data transmitted by the relay MT is fixed at $\rir=D_i$, according to Proposition \ref{proposition:concave}, the problem is convex with respect to $\pi_i$. Therefore, for this uni-variable convex optimization problem, the optimal solution can be obtained by checking the first-order condition of optimality. However, the objective function of problem $\mathrm{(P2')}$ is still complicated, for which the derivative is hard to obtain. Hence, we propose Algorithm I based on the derivative-free dichotomous search \cite{antoniou2007practical} to obtain the optimal solution numerically for problem  $\mathrm{(P2')}$.

\begin{table}[htp]
\begin{center}
\caption{One-dimensional dichotomous search algorithm for solving problem $\mathrm{(P2')}$ with precision $\delta_{\pi_i}$ and $\tau\ll1$.}
 \hrule\vspace{0.2cm} \textbf{Algorithm I}   \vspace{0.2cm}
\hrule \vspace{0.2cm}
\begin{itemize}
\item[1.] {\bf Initialize:} $\pi_i^{(l)} \coloneqq \epsilon$, $\pi_i^{(h)}\coloneqq \zeta_iE_i^{(D,S)}$, $\Delta_{\pi_i}\coloneqq|\pi_i^{(l)}-\pi_i^{(h)}|$;

\item[2.] {\bf Repeat:}
    \begin{itemize}
    \item[1.] Set temporary parameters:\\ $\tilde \pi_i^{(l)}\coloneqq \frac{1}{2}(\pi_i^{(l)}+\pi_i^{(h)})-\tau\Delta_{\pi_i}$,~$\tilde \pi_i^{(h)}\coloneqq\frac{1}{2}(\pi_i^{(l)}+\pi_i^{(h)})+\tau\Delta_{\pi_i}$;
    \item[2.] If $\mathbb{E}[C_i(\pi_i^{(l)})]<\mathbb{E}[C_i(\pi_i^{(h)})]$, set the price as $\pi_i^{(h)}\coloneqq\tilde \pi_i^{(l)}$;
    \item[3.] If $\mathbb{E}[C_i(\pi_i^{(l)})]>\mathbb{E}[C_i(\pi_i^{(h)})]$, set the price as $\pi_i^{(l)}\coloneqq\tilde \pi_i^{(h)}$;
    \item[4.] Otherwise, set $\pi_i^{(h)}\coloneqq\tilde \pi_i^{(h)}$ and $\pi_i^{(l)}\coloneqq\tilde \pi_i^{(l)}$;
    \item [5.] $\Delta_{\pi_i}\coloneqq|\pi_i^{(l)}-\pi_i^{(h)}|$;
    \end{itemize}
\item[3.] {\bf Until:} the condition $\Delta_{\pi_i}>\delta_{\pi_i}$ is violated;
\item[4.] ${\pi}_i^*\coloneqq(\pi_i^{(h)}+\pi_i^{(l)})/2$, ~~  $\mathbb{E}[{C_i^*}]\coloneqq\mathbb{E}[C_i({\pi}_i^*)]$.
\end{itemize}
\vspace{0.2cm} \hrule \label{algorithm:2}
\end{center}
\end{table}

\vspace{0.5em} 
\subsubsection{Joint Pricing and Load Sharing for Splittable Data}\label{subsec:wOLS}~\\

\vspace{-1em}
 Next, we discuss the general case where the data is splittable at the source MT in problem {$\mathrm{(P2)}$}. According to Proposition \ref{proposition:concave}, the objective function of problem {$\mathrm{(P2)}$} is convex with respect to $\pi_i$ given a fixed $\rir$ and to $\rir$ given a fixed $\pi_i$. Hence, based on the dichotomous search algorithm in Algorithm I, we propose Algorithm II that approximately minimizes the expected cost of the source MT $i\inm{K}_S$ with alternative optimization.

\begin{table}[htp]
\begin{center}
\caption{Alternative optimization algorithm for solving problem {$\mathrm{(P2)}$ with precision $\delta_{C_i}$}.}
 \hrule\vspace{0.2cm} \textbf{Algorithm II} \vspace{0.2cm}
\hrule \vspace{0.2cm}
\begin{itemize}
\item[1.] {\bf Initialize:} $n\coloneqq0$, $\rir\coloneqq D_i$, $\mathbb{E}[C_i^{(0)}]\coloneqq \zeta_iE_i^{(D,S)}$;
\item[2.] {\bf Repeat:}
    \begin{itemize}
    \item[1.] Optimize the  objective of problem {$\mathrm{(P2)}$} with respect to $\pi_i$ by dichotomous search with $\rir$ fixed ;
  	\item[2.] Optimize the  objective of problem {$\mathrm{(P2)}$} with respect to $\rir$ by dichotomous search with $\pi_i$ fixed ;
    \item [3.] $n\coloneqq n+1$;
    \end{itemize}
\item[3.] {\bf Until:} the condition $|\mathbb{E}[C_i^{(n)}]-\mathbb{E}[C_i^{(n-1)}]|>\delta_{C_i}$ is violated.
\end{itemize}
\vspace{0.2cm} \hrule \label{algorithm:2}
\end{center}
\end{table}

 For Algorithm II, it starts with the optimal solution obtained in Algorithm I with $D_i^{(R)}=D_i$. The algorithm then proceeds by iteratively optimizing and updating $\pi_i$ and $\rir$ with the other fixed until the stopping condition is satisfied. It should be noted that the algorithm always converges to a certain value within the range of $\delta_{C_i}$ from at least a locally optimal solution. This is because each iteration of the algorithm reduces the objective value and the optimal value of problem {$\mathrm{(P2)}$} is lower bounded. 

{Finally, for the complexity of Algorithm I, the the maximum number of iterations required for the searching of the optimal pricing ${\pi_i}$ with precision $\delta_{\pi_i}$ is $\cO(\log_2\frac{\zeta_iE_i^{(D,S)}}{\delta_{\pi_i}})$. Next, for the complexity of Algorithm II, the upper bound of each line search for $D_i^{(R)}$ and $\pi_i$ are $N_{D_i^{(R)}}=\log_2(\frac{D_i}{\delta_{D_i^{(R)}}})$ and $N_{\pi_i}=\log_2\frac{\zeta_iE_i^{(D,S)}}{\delta_{\pi_i}}$, respectively, where $\delta_{D_i^{(R)}}$ is precision requirement for the line search of $D_i^{(R)}$. The upper bound for the total number of iterations in the above alternative optimization is $M=\frac{\zeta_iE_i^{(D,S)}}{\delta_{C_i}}$. Hence, the upper bound for the complexity of Algorithm II is $\cO(M(N_{\pi_i}+N_{D_i^{(R)}}))$. Moreover, given the data rate of the source MT, user density, energy cost and channel condition, the optimal solution can be computed off-line and stored in a look-up table for practical implementation.}

\begin{table}[t]\caption{General simulation setup}
\centering
\begin{tabular}{ll}
\toprule
Simulation Parameters& Values\\
\cmidrule(r){1-1}\cmidrule(l){2-2}
 Noise power  & $\sigma^2=-110$ \rm{dBm}  \\
 Path-loss exponent & $\alpha=3.6$\\
 Reference distance & $r_0=10$ \rm{m}\\
 Path-loss at $r_0$ & $G_0=-70$ \rm{dB}\\
 Relay MT reservation utility& $\epsilon=0.2$\\
 Cost reduction threshold &$\gamma_i=1$\\
 Maximum battery level& $B_{\mathrm{max}}=100$ \rm{J}\\
 Maximum unit energy cost &$\zeta_{\mathrm{max}}=1$\\
 \bottomrule
\end{tabular}
\label{simulation_setup}
\end{table}

\section{Numerical Results}\label{sec:numerical_examples}
{In this section, we first show the convergence of the algorithm for the single source MT and examine its performance under different transmission schemes.  Then, the simulation of multiple source MTs is given to show their real-time operation under our proposed protocol in a single-cell system. The general simulation parameters are given in Table \ref{simulation_setup} and specific simulation setup and parameters for the cases of single source MT and multiple source MTs will be elaborated later in each subsection.}

\begin{table}[t]\caption{Simulation setup for single source MT}
\centering
\begin{tabular}{ll}
\toprule
Simulation Parameters& Values\\
\cmidrule(r){1-1}\cmidrule(l){2-2}
Distance from the source MT $i\inm{K}_S$ to BS\tablefootnote{The typical range of LTE in the urban environment is 1-5 \rm{km}.\cite{yi2012radio}} & $r_i=50$ \rm{m} \\
Short-term fading of this single source MT $i$ & $\eta_i=0.5$\\
Initial battery level of the source MT $i$ & $B_i=10~\mathrm{J}$\\
 \bottomrule
\end{tabular}
\label{simulation_setup_single}
\end{table}
\subsection{Single Source MT}\label{simulation1}
{In this subsection, we consider the simulation for single source MT. We first show the convergence of Algorithm II for the partial cooperation with splittable data rate and compare the convergent cost to that with non-splittable data rate by Algorithm I. Then, we show the simulation results for the expected cost of the single source MT versus battery levels under different schemes. The specific simulation parameters for this case of single source MT are given in Table \ref{simulation_setup_single}.}

\subsubsection{Convergence of Algorithm II for partial cooperation}
\begin{figure}[t]
  \centering
  \includegraphics[width=9cm]{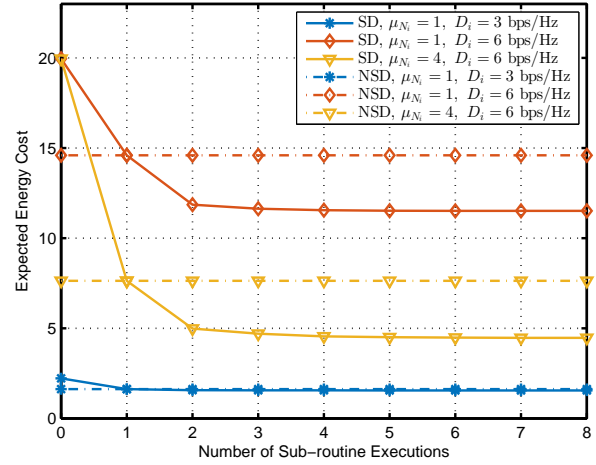}
  \caption{Expected energy cost of partial cooperation under incomplete information with splittable and non-splittable data  versus the number of sub-routine executions under different $\mu_{N_i}$'s and $D_i$'s.}\label{OptimalResult2}
\end{figure}

First, we show the convergence of Algorithm II for the partial cooperation with splittable data  compared with that with non-splittable data by Algorithm I in Section \ref{subsec:wOLS} for the data transmission of a single source MT $i\inm{K}_S$ under different data rates $D_i$ and average number of helping MTs $\mu_{N_i}$. First, exhaustive search on $\pi_i$ and $\rir$ with quantization of 0.2 and 0.1 in the feasible regions is conducted for three cases with different pairs of $\mu_{N_i}$ and $D_i$ and the minimum expected costs are  11.51, 4.46 and 1.55, respectively. Then, the result of the  joint optimization of $\pi_i$ and $D_i$ by Algorithm II with splittable data ({\it SD}) is shown in Fig. \ref{OptimalResult2} with the solid line.   The expected cost at iteration $\{0\}$ denotes the cost by the direct transmission, which are 19.96, 19.96 and 2.22, respectively.  The procedure in Algorithm II is executed 4 iterations and each of the uni-variable dichotomous search sub-routines in Algorithm II is executed 8 times, with sub-routines $\{1,3,5,7\}$ for minimization with respect to $\pi_i$   and sub-routines $\{2,4,6,8\}$ for that with respect to $\rir$ in Algorithm II. For comparison, the result by Algorithm I with non-splittable data ({\it NSD}) is shown by the three dash lines.

The simulation result is shown in Fig. \ref{OptimalResult2} and it can be observed that the convergence is fast.  The  converged expected energy costs for the three cases are 11.51, 4.46 and 1.55, respectively, which are the same as the results by exhaustive search. Hence, the global optimal solution is obtained in this case. Furthermore, the expected cost reduction from direct transmission for the three cases are 8.45, 15.50 and 0.67, respectively. Hence, according to the condition in (\ref{condition:CTmode}), the transmission mode selected by the source MT for the three cases will be CT, CT and DT, respectively. By comparing the two cases with $D_i=6~\mathrm{bps/Hz}$, it can be observed that a higher density of helping MTs can further reduce the expected energy cost.  By comparing the two cases with splittable and non-splittable data, it can also be observed that, in addition to the optimal pricing, load sharing can indeed further reduce the expected energy cost. Furthermore, the cost reductions with load sharing are 1.57, 1.25 and 1.12 times of those without load sharing, respectively. Therefore,  load sharing is more cost-effective when the size of the data is large and  the average number of helping MTs is small. {Finally, it should be noted that the case of splittable data leads to a lower energy cost compared with that of non-splittable data. This is because the energy consumption is exponentially increasing with respect to the transmission data rate and splitting the data and further optimizing the relay data rate result in a smaller total energy consumption.} 

In summary,  for the cooperative transmission under complete information with splittable data, the optimal solution is obtained by the following three-step procedure:  First, the source MT  computes the optimal data rate for each helping MT according to Proposition \ref{Prop:CompInfoNS}. Then, it searches for the one with the lowest energy cost from all the helping MTs by problem $\mathrm{(P1)}$. Last, it checks the condition in (\ref{compinfocondition}) and chooses between the DT and CT mode.

\begin{figure}[t]
  \centering
  \includegraphics[width=9cm]{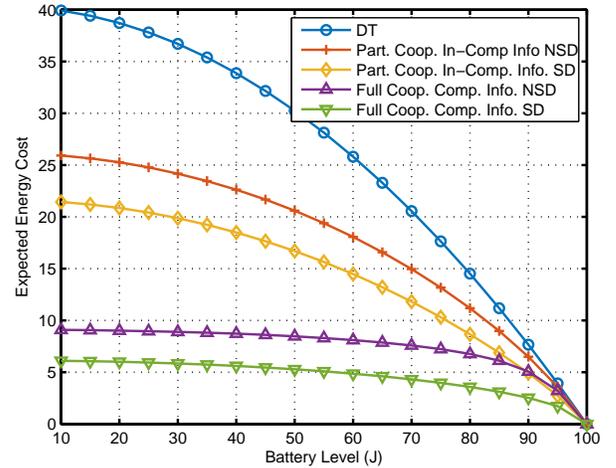}
  \caption{Expected cost of the single source MT versus its battery level under different schemes.}\label{expcost}
\end{figure}
\subsubsection{Expected energy cost under different battery levels and transmission schemes}
Next, we show the expected cost of different schemes under different battery levels. The simulation setup is shown as follows. We consider the simulation under  the following 5 schemes: 
\begin{itemize}
	\item {\bf Direct Transmission} {\it (DT)} in (\ref{directTrans}).
	\item {\bf Full Cooperation under Complete Information with Non-Splittable Data} {\it (Full Coop. Comp. Info. NSD)} in Section \ref{subsec:Incompnsd}.
	\item {\bf Full Cooperation under Complete Information with Splittable Data} {\it (Full Coop. Comp. Info. SD)} in Section \ref{subsec:compinfosd}.
	\item {\bf Partial Cooperation under Incomplete Information with Non-Splittable Data} {\it (Part. Coop. In-Comp. Info. NSD)} in Section \ref{subsec:woOLS}.
	\item {\bf Partial Cooperation under Incomplete Information with Splittable Data} {\it (Part. Coop. In-Comp. Info. SD)} in Section \ref{subsec:wOLS}.
\end{itemize}
Specifically, for the schemes of  partial cooperation under incomplete information with splittable and non-splittable data, the minimum expected costs are obtained by Algorithms I and II, respectively. For the schemes of  full cooperation under complete information with splittable and non-splittable data, the number of helping MTs $N_i$ for source MT $i\inm{K}_S$ is generated according to the  Poisson distribution with $\mu_{N_i}=2$ and the source MT transmits the data at the rate of $D_i=4\mathrm{~bps/Hz}$. Their battery levels $B_j,~j\inm{H}_i$ are uniformly generated on $[0,B_{\mathrm{max}}]$ and short term Rayleigh fading $\eta_j,~j\inm{H}_i$ is generated according to $\exp(1)$. The minimum energy costs can be obtained by the results in Sections \ref{subsec:Incompnsd} and \ref{subsec:compinfosd}, respectively, and the results are averaged over 1000 independent realizations for accurately obtaining the expected energy costs for comparison.  The expected cost of the transmission with DT mode is also obtained by averaging over 1000 independent realizations.

The simulation result is shown in Fig. \ref{expcost}. It can be observed that our proposed cooperative communications scheme performs significantly better than the direct transmission benchmark. Moreover, cooperative communications is more effective when the battery level of the source MT is low. This is  because when the battery level of the source MT is high and cost for direct transmission is low, it is less likely to seek help from the other MTs. Furthermore, it can also be observed that there are gaps in expected energy costs between the schemes with complete information in Section \ref{sec:completecoop} and those with incomplete information in Section \ref{sec:OptimalP2}, which are explained as follows:
\begin{itemize}
	\item [i)] In the case of complete information with non-splittable data, the source MT can observe the set of helping MTs as well as their channel conditions and battery levels, and choose the most cost-efficient one as the relay. While for the incomplete information case, the source MT can only randomly choose one from the possible helping MTs that accept the offer with the risk of ending up with direct transmission.
	\item [ii)] In the case of complete information with splittable data, in addition to the reason in i), the source MT can jointly optimize the relay data rate and the payment and choose the helping MT that leads to the minimum sum energy cost (as in Proposition \ref{Prop:CompInfoNS}). While in the case of incomplete information with splittable data, the source can only optimize the payment and relay data rate with respect to the expected energy cost, which has the possibility that the source MT ends up with direct transmission due to the lack of helping MTs' information. 
	\item [iii)] In both the cases with and without splittable data under complete information, the reservation utility margin $\epsilon$ for the helping MT is zero, which further reduces the cost  of the source MT from that under incomplete information and fully motivates the cooperation.   
\end{itemize}
{Finally, the figure shows that, when battery levels equal 100 J, the expected energy costs of all cases also become zero. This is due to our assumption that the energy cost at full battery capacity (i.e. battery level equals 100 J) is zero and at this case, there is no cooperation between the source and relay MTs.
}

\begin{figure}[t]
  \centering
  \includegraphics[width=8.8cm]{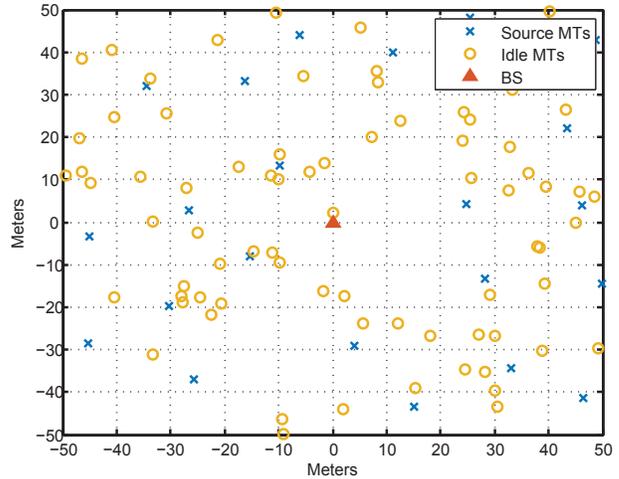}
  \caption{Setup for the simulation of multiple source MTs with $|\mathcal{K}|=100$ MTs.}\label{network}
\end{figure}

\begin{table}[t]\caption{Simulation setup for multiple source MTs}
\centering
\begin{tabular}{ll}
\toprule
Simulation Parameters& Values\\
\cmidrule(r){1-1}\cmidrule(l){2-2}
Total number of MTs & $|\mathcal{K}|=100$ \\
Probability at which MTs initiate data transmission & $\rho=0.2$\\
Normalized data rate\tablefootnote{The uplink spectrum efficiency of the LTE system is $3.75\sim 15~\mathrm{bps/Hz}$.\cite{LTEstandard}} & $D_i=6 ~\mathrm{bps/Hz}$\\
Range of the SRC for a source MT & $d=7~\mathrm{m}$\\
 \bottomrule
\end{tabular}
\label{simulation_setup_multiple}
\end{table}

\subsection{Multiple Source MTs}\label{SystemSimulation}
In this subsection, we  conduct a simulation with multiple source MTs and show the real-time operation of  our proposed cooperative communications protocol within a single cell. {We examine the five schemes considered in the previous subsection and show the performance improvement in terms of battery and communications outage, average battery level and battery level distribution, under a single-cell setup. The specific simulation parameters for  multiple source MTs are given in Table \ref{simulation_setup_multiple}  and the simulation setup is described as follows.}  

\begin{table*}[t]
\centering
\caption{Number of communications and battery outages for the five cases after 300 time slots.}\label{SimulationTable}
\begin{tabular}{lcC{1.3cm}C{1.3cm}C{1.3cm}C{1.3cm}}
\toprule
    & & \multicolumn{2}{c}{Part. Coop. In-Comp. Info.} & \multicolumn{2}{c}{Full Coop. Comp. Info.}\\
\cmidrule(r){3-4}\cmidrule(r){5-6}
    &DT  & \multicolumn{1}{c}{NSD} & \multicolumn{1}{c}{SD} & \multicolumn{1}{c}{NSD} & \multicolumn{1}{c}{SD}   \\
    \cmidrule(r){1-1}\cmidrule(r){2-2}\cmidrule(r){3-3}\cmidrule(r){4-4}\cmidrule(r){5-5}\cmidrule(r){6-6}
    Commun. Outage & 289     & 209    &  153    & 47 & 30  \\
    Battery Outage & 44     & 32    & 26    &   9  & 6  \\
    \bottomrule
    \vspace{0.3em}
\end{tabular}
\end{table*}

We consider our simulation within a $100\times100~ \mathrm{m}^2$  square area as shown in Fig. \ref{network}. The operation of the system begins with the battery levels of the MTs uniformly generated on $[0,B_{\mathrm{max}}]$.  For the purpose of investigation, at the beginning of each time slot, the positions of the MTs are uniformly re-generated within the above mentioned area. In this setup, it is possible that there is overlap between the set of helping MTs for different source MTs, where one helping MT can possibly be associated with two source MTs. In order to avoid this situation, we re-generate the positions of the source MTs if there is overlap between the helping MTs. According to the function $\mu_{N_i} =(1-\rho)\lambda \pi d^2,~i\inm{K}_S$, the average number of helping MTs in this setup is $\mu_{N_i}=1.2$. Due to the physical constraint of the MT, we set the maximum transmit energy of the MT as $E_{\mathrm{max}}=3~\mathrm{J}$ for any time slot. {If the transmit energy of the MT exceeds $E_{\mathrm{max}}$, a {\it communications outage} will be declared by the MT and the data package is discarded.}
During the operation of the system, if the battery of a certain MT is drained out, this MT declares a {\it battery outage} and ceases any operation from that time on, including data transmission as a source MT or cooperative relay for the other source MTs as relay MT.



We show the total number of  communications and battery outages for the 100 MTs after 300 time slots for the same 5 schemes as in Section \ref{simulation1}. The simulation results are shown in Table \ref{SimulationTable}. It can be observed that, compared with the benchmark case of DT,  all of the four proposed schemes  with cooperative communications perform better in terms of communications and battery outage. The reduction of the battery outages reflects the effectiveness of our protocol design for the energy saving of the MT, especially for those MTs that are low in battery level. In addition to the reduction of the battery outage, our proposed scheme also shows significant reduction in the number of  communications outage. This is because, in the case of direct transmission, if the channel condition of the source MT is poor, the transmission power will exceed the peak power constraint $E_{\mathrm{max}}$ and communications outage will occur. While, under the same circumstances with cooperative communications, the source MT can seek help from the other helping MTs, whose transmit power is possibly lower than the peak power constraint and the transmission can be successful. Hence, our proposed schemes can improve the uplink data transmission of the MTs in terms of both the communications reliability and battery sustainability.

\begin{figure}[t]
  \centering
  \includegraphics[width=9cm]{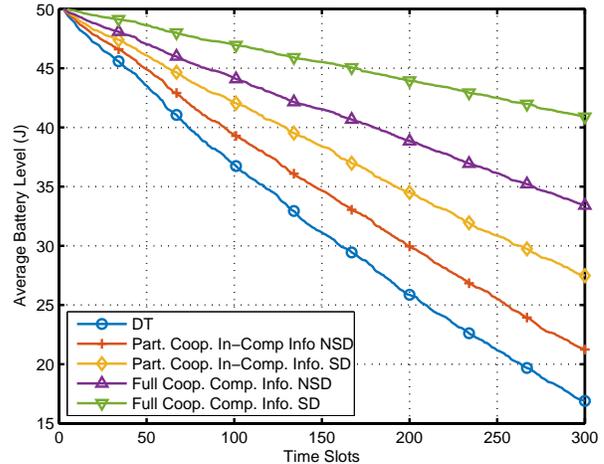}
  \caption{Average battery level $\sum_k{B_k}/|\mathcal{K}|$ of the MTs over time.}\label{avgbatterylevel}
\end{figure}
\begin{figure}[t]
  \centering
  \includegraphics[width=9cm]{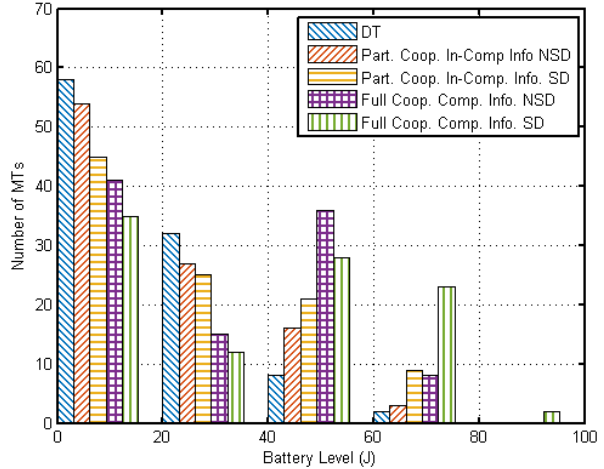}
  \caption{Distribution of the battery levels of 100 MTs after 300 time slots.}\label{BatteryLvlDistri}
\end{figure}

Next, we show the average battery level $\sum_k{B_k}/|\mathcal{K}|$ of the MTs  during the 300  time slots in Fig. \ref{avgbatterylevel}. It can be observed that the average battery levels of different schemes drop with different rates. Compared with the benchmark-case of DT, our proposed protocol can effectively increase the average battery level of the MTs over time. Even though the MTs under cooperative communications successfully deliver more data packages as shown in Table \ref{SimulationTable}, these schemes still perform better in terms of average battery level.

Finally, we show the distribution of the battery levels of the 100 MTs at the end of the 300 time slots in Fig. \ref{BatteryLvlDistri}.  It can be observed that, for the benchmark case of DT, a large proportion of the MTs have drained out their batteries. While for the other cases with cooperation, their battery levels remain on the relatively higher level than the direct transmission case by the distribution. It should also be noted that although a lot of MTs under cooperative communicationss stay in the low battery region (i.e. $\mathrm{0-20~J}$), their batteries are not empty according to Table \ref{SimulationTable}. This is because, when their battery levels are low, these MTs can possibly receive help from the other MTs such that their battery levels can be sustained. While, for the direct transmission case, the batteries of a lot of MTs in this region are empty due to the lack of help from the other MTs.

\section{Conclusion and Future Work}\label{sec:conclusion}
This paper studies the optimal pricing and load sharing for the energy saving of MTs with wireless  cooperative communications.  We formulate the MTs' decision making problem under uncertainties as an optimization problem for minimizing the expected energy cost of each source MT. The benchmark case of full cooperation under complete information is first considered for the cases of splittable and non-splittable data. Then, the general case of partial cooperation under incomplete information is considered and the optimal solutions are obtained by efficient dichotomous search and  alternative optimization algorithms.  Finally,  simulations with single source MT and multiple source MTs are given and show that our proposed cooperative communications protocol can significantly decrease the number of communications and battery outages  for the MTs and increase the average battery level during their operations. Overall, our results reveal new insights on the energy saving of the mutually beneficial cooperative communications, while hopefully lead to practical and energy-efficient design of wireless system with cooperative communications. 

In this paper, we consider single--relay selection to keep the communications overhead of the cooperation low. Clearly, a more general scenario is to consider multiple relay selection. Since the number of potential helping MTs is unknown to the source MT, the amount of data that each helping MT relays can only be determined by the helping MTs. In this case, the problem can be formulated as a Stackelberg game, where in the first phase, the source MT announces the amount of data to be relayed by the helping MTs and the payment by minimizing its own cost. Then, in the second phase, the helping MTs negotiate and compete with each other on the relaying data and payment by maximizing their own utilities. This two-stage game introduces time dynamics into the problem, which is challenging, while worth further investigation. 

\appendices

\section{Proof of Proposition \ref{Prop:CompInfoNS}}\label{appen:prop}
First, we take the first-order derivative of the sum energy cost $\zeta_jE_j^{(C,R)}+\zeta_iE_i^{(C,S)}$ with respect to $D_i^{(R)}$  and obtain
\begin{align}
-\ln 2\frac{\zeta_i\sigma^2}{G_i\eta_i}2^{D_i-D_i^{(R)}}+\ln 2\frac{\zeta_j\sigma^2}{G_i\eta_j}2^{D_i^{(R)}}.
\end{align}
Then, by setting it to zero, we obtain the stationary point of the objective function as
\begin{align}
 \tilde D_i^{(R)}=\frac{1}{2}\left(D_i+\log_2\frac{\theta_i}{\theta_j}\right).
 \end{align} 
Finally, we discuss the optimal solution in the region of $[0,D_i]$ for the following 3 cases.
 \begin{itemize}[leftmargin=*]
 	\item $\tilde D_i^{(R)}\in (-\infty,0]$: In this case, the sum energy cost is monotonically increasing within the region of $(0,D_i]$. Hence, the optimal solution is $\hat D_i^{(R)}=0$.
 	\item $\tilde D_i^{(R)}\in (0,D_i]$: The stationary point $\tilde D_i^{(R)}$ is contained  in this region. Hence, the optimal solution is $\hat D_i^{(R)}=\frac{1}{2}\left(D_i+\log_2\frac{\theta_i}{\theta_j}\right)$.
 	\item $\tilde D_i^{(R)}\in (D_i,\infty)$: In this case, the sum energy cost is monotonically decreasing within the region  $(0,D_i]$. Hence, the optimal solution is $\hat D_i^{(R)}=D_i$.
 \end{itemize}
The proof of Proposition 3.1 is thus completed.

\section{Proof of Proposition \ref{proposition:concave}}\label{Sec:Appen}
We need to prove the objective of the problem $\mathrm{(P2)}$ is marginally convex with respect to  $\pi_i$ and $\rir$.  Denote \begin{align}
\psi_{I}=1-\left(1-\frac{w_i}{\zeta_{\mathrm{max}}}\left(1-e^{-\frac{\zeta_{\mathrm{max}}}{w_i}}\right)\right)^{n}\nonumber,
\end{align}
where  $w_i=\frac{G_i(\pi_i-\epsilon)}{\sigma^2(2^{\rir}-1)}$
 and 
\begin{align}
 \psi_{II}= \pi_i+\zeta_i\eis- \zeta_iE_i^{(D,S)}\nonumber.
\end{align}
We first prove that $\psi_{I}\psi_{II}$ is marginally convex with respect to $\pi_i$ and $\rir$. Then,  the convexity of problem $\mathrm{(P2)}$'s objecive function  
\begin{align}
\sum_{n=0}^{\infty}\mathrm{Pr}(N_i=n) \left\{\psi_{I}\psi_{II}+\zeta_iE_i^{(D,S)}\right\}\nonumber
\end{align}
follows naturally. 

\begin{itemize}[leftmargin=*]
\item First, we  prove that the  $\psi_{I}\psi_{II}$ is marginally convex  with respect to $\pi_i$. In the first place, we denote $\psi_{I}(\pi_i)=1-\psi_j(\pi_i)^{N_i},~j\inm{H}_i$ and prove that $\psi_j(\pi_i)=1-\frac{w_i}{\zeta_{\mathrm{max}}}\bigg(1-e^{-\frac{\zeta_{\mathrm{max}}}{w_i}}\bigg)$ is a convex function with respect to $\pi_i$. By taking the second-order derivative of  $\psi_j(\pi_i)$, we can have
 \begin{align}
 \psi_j''(\pi_i)=\frac{\zeta_{\mathrm{max}} e^{-\frac{\zeta_{\mathrm{max}}}{w_i}}}{w_i(\pi_i-\epsilon)^2},~j\in\mathcal{H}_i,~i\in\mathcal{K}_S.
 \end{align}
Since $\psi_j''(\pi_i)>0$, $\psi_j(\pi_i)$ is a convex function. It can be verified that $\psi_j(\pi_i)^{N_i}$ is also convex because $\psi_j(\pi_i)$ is monotonically decreasing and convex. Hence, $\psi_{I}(\pi_i)$ is concave and monotonically increasing.

\hspace{1em}Then, it can be verified that $\psi_{II}''(\pi_i)=0$ since $\psi_{II}(\pi_i)$ is a linear function. Because $\psi_{I}(\pi_i)$ and $\psi_{II}(\pi_i)$ are both monotonically increasing with respect to $\pi_i$, it can be obtained that $\psi_{I}'(\pi_i)\psi_{II}'(\pi_i)>0$. Also, due to the fact that  $\psi_{I}''(\pi_i)<0$ and $\psi_{II}(\pi_i)\leq 0$, $\psi_{I}''(\pi_i)\psi_{II}(\pi_i)$ is also positive.  Then, the second-order derivative of $\psi_{I}(\pi_i)\psi_{II}(\pi_i)$ can be expressed as
\begin{align}\label{2ndOrderDerivative}
&\underbrace{\psi_{I}''(\pi_i)}_{\leq 0}\underbrace{\psi_{II}(\pi_i)}_{\leq 0}+2\underbrace{\psi_{I}'(\pi_i)}_{\geq 0}\underbrace{\psi_{II}'(\pi_i)}_{\geq 0}+\psi_{I}(\pi_i)\underbrace{\psi_{II}''(\pi_i)}_{=0}.
\end{align}
Therefore, it can be verified that (\ref{2ndOrderDerivative}) is positive and thus  $\psi_{I}(\pi_i)\psi_{II}(\pi_i)$ is a convex function with respect to $\pi_i$ with $\rir$ fixed. 

\item Second, by a similar approach as above, we can  prove that $\psi_{I}(\rir)$ is monotaonically decreasing and concave with respect to  $\rir$  and $\psi_{II}(\rir)$ is  montonically decreasing and convex with respect to  $\rir$. Then, by taking the second-order derivative of $\psi_{I}(\rir)\psi_{II}(\rir)$ with respect to $D_i^{(R)}$, we have
\begin{align}
&\underbrace{\psi_{I}''(\rir)}_{\leq 0}\underbrace{\psi_{II}(\rir)}_{\leq 0}+2\underbrace{\psi_{I}'(\rir)}_{\leq 0}\underbrace{\psi_{II}'(\rir)}_{\leq 0}\nonumber\\
&+\underbrace{\psi_{I}(\rir)}_{\geq 0}\underbrace{\psi_{II}''(\rir)}_{\geq 0}.
\end{align}

Hence, we have proved that the  function $\psi_{I}(\rir)\psi_{II}(\rir)$ is also marginally convex with respect to $\rir$ with $\pi_i$  fixed. 
\end{itemize}

Proposition \ref{proposition:concave} thus follows.

\bibliographystyle{IEEEtran}
\bibliography{work2_journal}

\begin{thebibliography}{10}
\providecommand{\url}[1]{#1}
\csname url@samestyle\endcsname
\providecommand{\newblock}{\relax}
\providecommand{\bibinfo}[2]{#2}
\providecommand{\BIBentrySTDinterwordspacing}{\spaceskip=0pt\relax}
\providecommand{\BIBentryALTinterwordstretchfactor}{4}
\providecommand{\BIBentryALTinterwordspacing}{\spaceskip=\fontdimen2\font plus
\BIBentryALTinterwordstretchfactor\fontdimen3\font minus
  \fontdimen4\font\relax}
\providecommand{\BIBforeignlanguage}[2]{{%
\expandafter\ifx\csname l@#1\endcsname\relax
\typeout{** WARNING: IEEEtran.bst: No hyphenation pattern has been}%
\typeout{** loaded for the language `#1'. Using the pattern for}%
\typeout{** the default language instead.}%
\else
\language=\csname l@#1\endcsname
\fi
#2}}
\providecommand{\BIBdecl}{\relax}
\BIBdecl

\bibitem{CNN2005}
\BIBentryALTinterwordspacing
{CNN}, ``Battery life concerns mobile users.'' [Online]. Available:
  \url{http://edition.cnn.com/2005/TECH/ptech/09/22/phone.study/}
\BIBentrySTDinterwordspacing

\bibitem{Perrucci2011}
G.~P. Perrucci, F.~H.~P. Fitzek, and J.~Widmer, ``Survey on energy consumption
  entities on the smartphone platform,'' in \emph{Proc. IEEE 73rd Veh. Tech.
  Conf. (VTC Spring)}, May 2011, pp. 1--6.

\bibitem{Pathak2012}
A.~Pathak, Y.~C. Hu, and M.~Zhang, ``Where is the energy spent inside my {APP}?
  {Fine} grained energy accounting on smartphones with {Eprof},'' in
  \emph{Proc. of the 7th ACM Euro. Conf. on Computer Sys. (EuroSys '12)}, 2012,
  pp. 29--42.

\bibitem{kramer2007cooperative}
G.~Kramer, I.~Mari{\'c}, and R.~Yates, \emph{Cooperative Communications}, ser.
  Foundations and trends in networking.\hskip 1em plus 0.5em minus 0.4em\relax
  Now Publishers, 2007.

\bibitem{CuiGoldsmith2005}
S.~Cui, A.~Goldsmith, and A.~Bahai, ``Energy-constrained modulation
  optimization,'' \emph{IEEE Trans. Wireless Commun.}, vol.~4, no.~5, pp.
  2349--2360, Sep. 2005.

\bibitem{FuKim2011}
L.~Fu, H.~Kim, J.~Huang, S.-C. Liew, and M.~Chiang, ``Energy conservation and
  interference mitigation: From decoupling property to win-win strategy,''
  \emph{IEEE Trans. Wireless Commun.}, vol.~10, no.~11, pp. 3943--3955, Nov.
  2011.

\bibitem{KimCeciana2010}
H.~Kim and G.~de~Veciana, ``Leveraging dynamic spare capacity in wireless
  systems to conserve mobile terminals' energy,'' \emph{IEEE/ACM Trans.
  Networking}, vol.~18, no.~3, pp. 802--815, Jun. 2010.

\bibitem{LuoZhangLim2014}
S.~Luo, R.~Zhang, and T.~J. Lim, ``Joint transmitter and receiver energy
  minimization in multiuser {OFDM} systems,'' \emph{IEEE Trans. Commun.},
  vol.~62, no.~10, pp. 3504--3516, Oct 2014.

\bibitem{LuoZhangLim2015}
------, ``Downlink and uplink energy minimization through user association and
  beamforming in {C-RAN},'' \emph{IEEE Trans. Wireless Commun.}, vol.~14,
  no.~1, pp. 494--508, Jan. 2015.

\bibitem{ZhouCui2008}
Z.~Zhou, S.~Zhou, J.-H. Cui, and S.~Cui, ``Energy-efficient cooperative
  communication based on power control and selective single-relay in wireless
  sensor networks,'' \emph{IEEE Trans. Wireless Commun.}, vol.~7, no.~8, pp.
  3066--3078, Aug. 2008.

\bibitem{ZouZhuZhang2013}
Y.~Zou, J.~Zhu, and R.~Zhang, ``Exploiting network cooperation in green
  wireless communication,'' \emph{IEEE Trans. Commun.}, vol.~61, no.~3, pp.
  999--1010, Mar. 2013.

\bibitem{LiuWangGuo}
D.~Liu, W.~Wang, and W.~Guo, ``Green cooperative spectrum sharing
  communication,'' \emph{IEEE Commun. Let.}, vol.~17, no.~3, pp. 459--462, Mar.
  2013.

\bibitem{ReviewReco}
G.~Botter, J.~Alonso-Zarate, L.~Alonso, F.~Granelli, and C.~Verikoukis,
  ``Extending the lifetime of {M2M} wireless networks through cooperation,'' in
  \emph{IEEE Int'l Conf. Commun. (ICC)}, Jun. 2012, pp. 6003--6007.

\bibitem{YangFang2012}
D.~Yang, X.~Fang, and G.~Xue, ``Game theory in cooperative communications,''
  \emph{IEEE Wireless Commun.}, vol.~19, no.~2, pp. 44--49, Apr. 2012.

\bibitem{VirtualCurrency}
J.-P. Hubaux, T.~Gross, J.-Y. Le~Boudec, and M.~Vetterli, ``Toward
  self-organized mobile ad hoc networks: the terminodes project,'' \emph{IEEE
  Commun. Mag.}, vol.~39, no.~1, pp. 118--124, Jan. 2001.

\bibitem{WangHan2009}
B.~Wang, Z.~Han, and K.~Liu, ``Distributed relay selection and power control
  for multiuser cooperative communication networks using {Stackelberg} game,''
  \emph{IEEE Trans. Mob. Comput.}, vol.~8, no.~7, pp. 975--990, Jul. 2009.

\bibitem{YangHuang2013}
Y.~Yan, J.~Huang, and J.~Wang, ``Dynamic bargaining for relay-based cooperative
  spectrum sharing,'' \emph{IEEE J. Sel. Areas in Commun.}, vol.~31, no.~8, pp.
  1480--1493, Aug. 2013.

\bibitem{PIMRC}
H.-T. Lin, Y.-Y. Lin, and W.-C. Chang, ``Reputation auction framework for
  cooperative communications in green wireless networks,'' in \emph{IEEE 23rd
  Int'l Sym. Personal Indoor and Mob. Radio Commun. (PIMRC),}, Sep. 2012, pp.
  875--880.

\bibitem{KandeepanJayaweera2012}
S.~Kandeepan, S.~Jayaweera, and R.~Fedrizzi, ``{Power-Trading} in wireless
  communications: A cooperative networking business model,'' \emph{IEEE Trans.
  Wireless Commun.}, vol.~11, no.~5, pp. 1872--1880, May 2012.

\bibitem{StochasticGeometry}
M.~Haenggi, J.~Andrews, F.~Baccelli, O.~Dousse, and M.~Franceschetti,
  ``Stochastic geometry and random graphs for the analysis and design of
  wireless networks,'' \emph{IEEE J. Sel. Areas Commun.}, vol.~27, no.~7, Sep.
  2009.

\bibitem{Kingman1993}
J.~Kingman, \emph{Poisson Process}.\hskip 1em plus 0.5em minus 0.4em\relax
  Oxford University Press, 1993.

\bibitem{WifiDirect}
\BIBentryALTinterwordspacing
{WiFi Alliance}, ``{Discover Wi-Fi Wi-Fi Direct},'' 2013. [Online]. Available:
  \url{http://www.wi-fi.org/discover-wi-fi/wi-fi-direct}
\BIBentrySTDinterwordspacing

\bibitem{bluetooth}
\BIBentryALTinterwordspacing
{Bluetooth SIG Inc.}, ``Bluetooth low energy, bluetooth development portal,''
  2014. [Online]. Available:
  \url{https://developer.bluetooth.org/TechnologyOverview/Pages/BLE.aspx}
\BIBentrySTDinterwordspacing

\bibitem{LTEstandard}
\BIBentryALTinterwordspacing
T.~Nakamura, ``Proposal for candidate radio interface technologies for
  {IMT‐-Advanced} based on {LTE} {Release} 10 and beyond
  ({LTE‐-Advanced}),'' 2009. [Online]. Available:
  \url{http://www.3gpp.org/IMG/pdf/2009_10_3gpp_IMT.pdf}
\BIBentrySTDinterwordspacing

\bibitem{mas1995}
A.~Mas-Colell, M.~Whinston, and J.~Green, \emph{Microeconomic Theory}.\hskip
  1em plus 0.5em minus 0.4em\relax Oxford University Press, 1995.

\bibitem{fishburn1970utility}
P.~Fishburn, \emph{Utility theory for decision making}, ser. Publications in
  operations research.\hskip 1em plus 0.5em minus 0.4em\relax Wiley, 1970.

\bibitem{antoniou2007practical}
A.~Antoniou and W.-S. Lu, \emph{Practical optimization: algorithms and
  engineering applications}.\hskip 1em plus 0.5em minus 0.4em\relax Springer,
  2007.

\bibitem{yi2012radio}
S.~Yi, S.~Chun, Y.~Lee, S.~Park, and S.~Jung, \emph{Radio Protocols for {LTE}
  and {LTE-advanced}}.\hskip 1em plus 0.5em minus 0.4em\relax John Wiley \&
  Sons, 2012.

\end{thebibliography}

\end{document}